\documentclass{article}
\usepackage[preprint]{spconf}
\copyrightnotice{\copyright 2019 IEEE. DOI: {10.1109/ICME.2019.00142}}
\usepackage{amsmath,amssymb,epsfig}
\usepackage{hyperref}
\usepackage[utf8]{inputenc}
\usepackage{pgfplots, pgfplotstable}
\pgfplotsset{compat=newest}

\begin{document}\sloppy
\onecolumn

\section*{Preprint Notice}
This is the preprint of 
\\
\\
\textbf{Two-Staged Acoustic Modeling Adaption for Robust Speech Recognition by the Example of German Oral History Interviews}, Digital Object Identifier (DOI)  \href{https://doi.org/10.1109/ICME.2019.00142}{10.1109/ICME.2019.00142},
\\
\\
accepted for IEEE International Conference on Multimedia and Expo (ICME), Shanghai, China, July 2019.
\\
\\
\textcopyright\ 2019 IEEE. Personal use of this material is permitted. Permission from IEEE must be obtained for all other uses, in any current or future media, including reprinting/republishing this material for advertising or promotional purposes, creating new collective works, for resale or redistribution to servers or lists, or reuse of any copyrighted component of this work in other works.

\twocolumn
\pagestyle{empty}

\def\x{{\mathbf x}}
\def\L{{\cal L}}

\title{Two-Staged Acoustic Modeling Adaption for Robust Speech Recognition by the Example of German Oral History Interviews}

\name{Michael Gref$^{~1,2}$, Christoph Schmidt$^1$, Sven Behnke$^{1,3}$, Joachim Köhler$^1$}
\address{
  $^1$Fraunhofer Institute for Intelligent Analysis and Information Systems (IAIS), Germany\\
  $^2$Institute for Pattern Recognition (iPattern), Niederrhein Univ. of Applied Sciences, Germany\\
  $^3$Autonomous Intelligent Systems (AIS), Computer Science Institute VI, Univ. of Bonn, Germany\\
  \{michael.gref, christoph.andreas.schmidt, sven.behnke, joachim.koehler\}@iais.fraunhofer.de
	}

\maketitle

\begin{abstract}
In automatic speech recognition, often little training data is available for specific challenging tasks, but training of state-of-the-art automatic speech recognition systems requires large amounts of annotated speech. To address this issue, we propose a two-staged approach to acoustic modeling that combines noise and reverberation data augmentation with transfer learning to robustly address challenges such as difficult acoustic recording conditions, spontaneous speech, and speech of elderly people. We evaluate our approach using the example of German oral history interviews, where a relative average reduction of the word error rate by $19.3 \%$ is achieved.
\end{abstract}

\begin{keywords}
Robust speech recognition, domain adaption, transfer learning, multi-condition training, data augmentation, oral history
\end{keywords}

\section{Introduction}
Automatic speech recognition (ASR) has undergone enormous improvements in recent years. Nowadays, it is successfully used in many applications, both in the commercial and industrial sectors. ASR not only enables the development of smart speech assistants but is also used for subtitling, information mining, analytics, and recommendation. 

However, training state-of-the-art ASR systems requires large amounts of annotated speech. If training data is not available to a sufficient extent, only unsatisfactory results are achieved. Especially for challenging scenarios, often only little training data is available, and off-the-shelf ASR systems perform poorly. Such challenges can arise from different acoustic conditions such as noise and reverberation, but also varying recording equipment, spontaneous, fast speech, unclear pronunciations and dialects can be very challenging. 

In this work, we propose an approach to tackle these challenges by combining multi-condition training via data augmentation and transfer learning on very little data in a two-staged acoustic modeling adaption. We evaluate our approach using the example of German oral history interviews, in which all aforementioned challenges occur to varying degrees.

\section{Related Work}

\label{sec:intro}
\begin{figure}
	\centering
	\centerline{\epsfig{figure=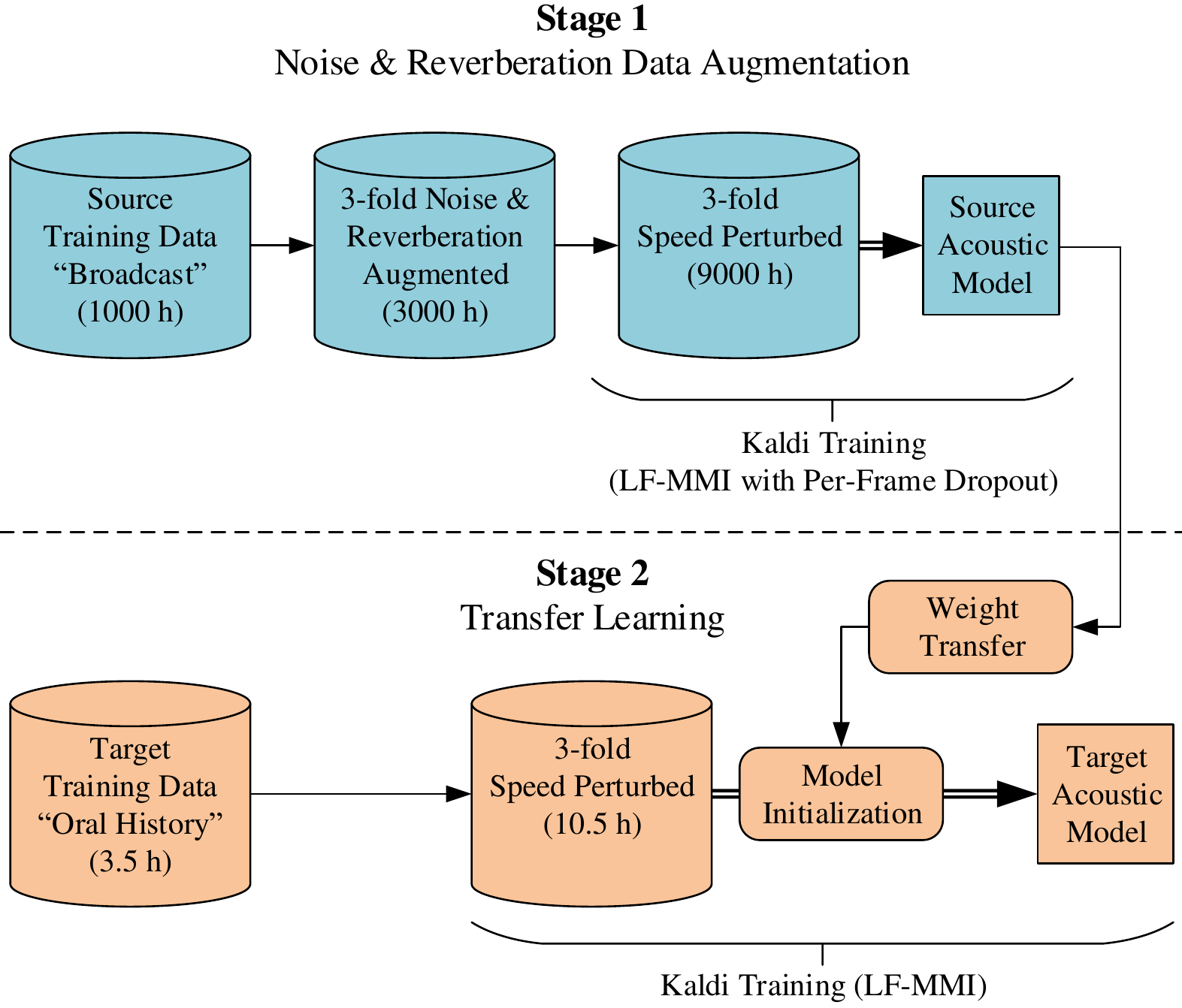,width=8.5cm}}
	\caption{Proposed approach. Noise and reverberation data augmentation is applied in Stage 1 to obtain a robust acoustic source model. In Stage 2, transfer learning is applied to tackle further challenges such as spontaneous speech.}
	\label{fig:teaser_pic}
\end{figure}

ASR is a popular and highly researched area and new approaches are regularly proposed. Currently, lattice-free maximum mutual information (LF-MMI) trained models achieve state-of-the-art results on many different ASR tasks \cite{Povey.2016.LF_MMI}.  

Oral history in historical research refers to conducting and analyzing interviews with contemporary witnesses. In Germany, this kind of research focused above all on the period of the Second World War and National Socialism. In the meantime, it has also come to include many other topics and historical periods. In our prior work \cite{Gref.2018.oralhistory01, Gref.2018.oralhistory02} we studied the application and adaption of state-of-the-art ASR to German oral history interviews.

Applying data augmentation to training data is a common approach to increase the amount of training data in order to improve the robustness of a model. In ASR it can be used, e.g., to apply multi-condition training, when no real data in the desired condition is available. Data augmentation is, however, limited to acoustic effects that can be created in a sufficiently realistic manner - such as additive noise and reverberation. The data augmentation of reverberant speech for state-of-the-art LF-MMI models has been studied by Ko et al.\ \cite{Ko.2017.KaldiReverbDataAugmentation}. Several speed perturbation techniques to increase the training data variance have been investigated by Ko et al.\ \cite{Ko.2015.SpeedPerturbation}. The proposed method in this work is to increase the data three-fold by creating two additional versions of each signal using the constant speed factors $ 0.9 $ and $ 1.1 $---a method that is used in many recent Kaldi training routines by default.

Transfer learning is an approach used to transfer knowledge of a model trained in one scenario to train a model in another related scenario to improve generalization and performance \cite{Goodfellow.2016.DeepLearningBook}. It is particularly useful in scenarios where only little training data is available for the main task but a large amount of annotated speech is available for a similar or related task. A detailed overview of transfer learning in speech and language processing is given by Wang et al.\ \cite{Wang.2015.transfer_learning_overview_in_speechprocessing}. Transfer learning for ASR systems using LF-MMI models has been studied by Ghahremani et al.\ \cite{Ghahremani.2017.KaldiTransferlearning_LF_MMI} for many different common English speech recognition tasks. 

However, most works in ASR, such as the aforementioned, studied transfer learning with a much greater amount of annotated speech than is available in the oral history task, for instance. In addition, most works focus on either data augmentation or transfer learning, usually to address a particular task or challenge, such as robustness to noise and not the robustness of an acoustic model as a whole.

\section{Proposed Approach}
We aim at improving the performance of robust ASR systems by performing a two-staged acoustic modeling adaption using a very little amount of target training data. An overview of the proposed method is given in Fig.\ \ref{fig:teaser_pic}. In the first stage, multi-condition training is applied using noise and reverberation data augmentation to obtain a robust acoustic source model. The second stage applies transfer learning to tackle the remaining challenges of the target data that could not be synthesized in the first stage, such as spontaneous speech, dialectics and pronunciations.

The first stage of the approach is based on our prior work \cite{Gref.2018.oralhistory02}, where multi-condition training using noise and reverberation data augmentation was used to decrease the acoustic mismatch of conventional clean training data and oral history interviews. This has been proven to significantly increase the performance of ASR systems on German oral history interviews. In contrast to the aforementioned work, where the amount of training data is kept to the same size, in our approach the data is increased 3-fold.

Defining discrete-time-signals as sequences of sample values, the applied augmentation can be described as
\begin{equation}
\label{eqn:data_aug_with_noise}
({x}_n)_{n \in \mathbb{N} } := (s_n)_{n \in \mathbb{N} } * (h_n)_{n \in \mathbb{N}} + (w_n)_{n \in \mathbb{N} } * (\tilde{h}_n)_{n \in \mathbb{N} }
\end{equation}

if both noise and reverberation inside a simulated room affects the speech signal. Here, $ * $ is the convolution operation for sequences, $ (s_n)_{n \in \mathbb{N} } $ the sequence of the clean speech signal, $ (h_n)_{n \in \mathbb{N} } $, $ (\tilde{h}_n)_{n \in \mathbb{N} } $ are room impulse responses modeling the reverberation of one room at different positions and $ (w_n)_{n \in \mathbb{N} } $ describes the sequence of the noise signal. If only reverberation and no background noise affects the speech signal, $ \forall n \in \mathbb{N}: w_n = 0 $ applies and yields 
\begin{equation}
\label{eqn:data_aug_without_noise}
(x_n)_{n \in \mathbb{N} } := (s_n)_{n \in \mathbb{N} } * (h_n)_{n \in \mathbb{N}}. 
\end{equation}

We use $ 266 $ room impulse responses of small and me\-di\-um sized rooms along with $ 14.5 $ hours of real-life noise recordings collected from different sources - such as the Aachen Impulse Response database \cite{Jeub.2009.AachenRirDB}, other freely available and in-house data. We create the following two artificially corrupted versions of our source training data and merge them with the original (clean) set to create a $ 3000 $ hour multi-condition source training set: 

\begin{itemize}
	\item \textbf{Reverb}: All signals are convolved according to Equation (\ref{eqn:data_aug_without_noise}) with randomly selected room impulse responses of small or medium sized rooms. No noise is applied here.
	
	\item \textbf{Reverb+RealNoise}: Similar to \textbf{Reverb} but added noise recordings according to equation (\ref{eqn:data_aug_with_noise}) applying a random signal-to-noise ratio between $ 10 $ and $ 20 $ dB. The noises have been randomly selected from real-life recordings, e.g.\ street noises, bus noises, police sirens, hairdryers. To avoid overfitting, we randomly selected and superposed up to three different noises for one audio file before applying the reverberation.
\end{itemize}

The transfer learning in Stage 2 is inspired by the work of Ghahremani et al.\ \cite{Ghahremani.2017.KaldiTransferlearning_LF_MMI}. In our setup, a full weight transfer of the entire source model for initialization of the target model is applied without any layer freezing. In particular, the output layer is not replaced in contrast to some other transfer learning approaches in ASR, since the same set of phonemes and the same decision tree is used both in the source and target scenario. In the transfer learning stage, the i-vector extractor of the model trained in Stage 1 is used without any adaption.

The neural network training routine in Stage 2 is almost equal to the one used in Stage 1 with only slight adjustments. An overview of the parameters that are different in the transfer learning stage is given in Table \ref{tab:training_paramters}. In Stage 1 we apply per-frame dropout according to Cheng et al.\ \cite{Cheng.2017.KaldiLSTMDropout}. The training in Stage 2 is performed without dropout. Our previous experiments with transfer learning showed that dropout seems to reduce the performance training on small data sets. In both stages, the training is performed for four epochs with a reducing learning rate. The initial and final learning rate in the second stage is lower than in the first stage due to the significantly smaller amount of training data. Note that 3-fold speed perturbation is applied in every setup, since we consider this technique to be a default procedure in the Kaldi training routines. 

\begin{table}[t]
	\begin{center}
		\caption{Changed training parameters in Stage 1 and 2} \label{tab:training_paramters}
		\begin{tabular}{|c|r|r|}
			\hline
			\textbf{Parameter} 				& \textbf{Stage 1} 			& \textbf{Stage 2}	\\
			\hline
			Init./final learn rate	&  1e-3 / 1e-4		& 1e-6 / 1e-7  \\
			Dropout Schedule		& 0, 0@0.2, 0.3@0.5, 0 & 0, 0 \\
			\hline
		\end{tabular}
	\end{center}
\end{table}

\section{Experimental Setup}

\subsection{Lexicon and language model}
The lexicons needed for training in Stage 1, Stage 2 and decoding are all obtained using the same grapheme-to-phoneme pronunciation model trained with Sequitur G2P \cite{Bisani.2008.SequiturG2P}. This model is created using the German pronunciation database Phonolex from the Bavarian Archive for Speech Signals.

For decoding, we use a 500,000 words 5-gram broadcast language model. This model is trained on broadcast text corpora consisting of $ 75 $ million words. Decoding parameters are kept the same for all experiments. In particular, the language model weight is kept to a fixed value for all experiments.

\subsection{Acoustic model}

\subsubsection{Training data}
For training the source system in the first stage, we utilize a 1000 hour large-scale corpus of German broadcast speech data \textit{GerTV1000h} \cite{Stadtschnitzer.2014.GerTV1000hCorpus}. This data set can be considered to be out of domain for the oral history scenario, since the broadcast recordings differ from oral history in terms of the used recording technology, audio signal quality and speech characteristics.

As target data, we use the oral history data set proposed in our prior work \cite{Gref.2018.oralhistory01}. It consists of $ 3.5 $ hours audio from $ 35 $ different speakers recorded in real oral history interviews. All audio signals are resampled to the sample frequency of the training data ($ 16 $ kHz). The set contains 27,708 transcribed spoken words with a vocabulary of 4582 words. The recordings took place between 1980 and 2012, representing a wide range of recording technology, interview methodology, dialects and pronunciations. The set is manually transcribed and segmented and has an average segment length of $ 5.3 $ seconds with overall $ 2392 $ segments. 

\subsubsection{Acoustic model neural network topology}
All acoustic model networks use a $ 300 $-dimensional input at each time-step consisting of five consecutive $ 40 $-dimensional MFCC features and a $ 100 $-dimensional i-vector \cite{Dehak.2011.ivector}. We use a topology with ten hidden layers that was proposed and investigated by Cheng et al.\ \cite{Cheng.2017.KaldiLSTMDropout}. The acoustic model neural network consists of seven TDNN layers \cite{Waibel.1989.TDNN, Peddinti.2015.KaldiTDNN} and three LSTM layers \cite{Hochreiter.1997.LSTM} stacked in the order given in Table \ref{tab:am_nnet_topology}.

The applied implementation uses LSTM layers with forget gates \cite{Gers.2000.LSTMLearningToForget}, peephole connections \cite{Gers.2000.LSTMPeepholes} and projection layers \cite{Sak.2014.LSTMProjection}. The LSTM layers have a cell dimension of $ 1024 $ and a projection dimension of $ 256 $. The TDNN layers are $ 1024 $-dimensional.

\begin{table}
	\begin{center}
		\caption{Hidden Layer of the Acoustic Model} \label{tab:am_nnet_topology}
		\begin{tabular}{|r|c|c|}
			\hline
			\textbf{\#} & \textbf{Type} & \textbf{Temporal Context}	\\
			\hline
			1 & TDNN 	& $ \{ -2, -1, 0, 1, 2 \} $ \\
			2 \& 3 & TDNN 	& $ \{ -1, 0, 1 \} $ \\
			4 & LSTM 	& $ - $ \\
			5 \& 6 & TDNN 	& $ \{ -3, 0, 3 \} $ \\
			7 & LSTM 	& $ - $ \\
			8 \& 9 & TDNN 	& $ \{ -3, 0, 3 \} $ \\
			10 & LSTM 	& $ - $ \\
			\hline
		\end{tabular}
	\end{center}
\end{table}

\subsubsection{Experiments}
All experiments are carried out using the Kaldi ASR toolkit \cite{Povey.2011.KaldiToolkit}. As part of the Kaldi training routines, the aforementioned speed perturbation \cite{Ko.2015.SpeedPerturbation} is applied on the entire training data to increase the amount of data three-fold before neural network training. All models are trained using the LF-MMI \cite{Povey.2016.LF_MMI} criterion. Overall, four major types of setups are examined in our experiments:

\begin{enumerate}
	\item \textbf{Baseline}: For comparison, we train a baseline acoustic model with the same setup as in Stage 1, excluding the noise and reverberation data augmentation and the entire transfer learning Stage 2. 
	
	\item \textbf{Stage 1 Only (Data Augmentation)}: Evaluating the performance of the source model trained in Stage 1 using the noise and reverberation data augmentation on the $ 35 $ speaker sets.
	
	\item \textbf{Stage 2 Only (Transfer Learning)}: Applying the transfer learning experiments on the clean-trained baseline model.

	\item \textbf{Proposed Approach}: Applying Stage 1 and Stage 2.
\end{enumerate}

\subsubsection{Leave-one-speaker-out evaluation}
Since only very little data from the target domain is available, we apply a leave-one-speaker-out evaluation on the target data. This approach can be understood as a $ k $-fold cross-validation where the data set is partitioned according to speakers. This means each subset consists of exactly one speaker. Then we loop over the data subsets and keep one speaker out of the training set for validation and train one model on the data of the remaining $ k-1 $ speakers. This way, we run $ k $ experiments in Stage 2 and evaluate each trained model on the speaker that was not present in the training data. We trained one model in Stage 1 and then used this model as the source model for all $ 35 $ different leave-one-speaker-out experiments in Stage 2.

\section{Results and Discussion}

\begin{table*}[t]
	\begin{center}
		\caption{Word error rates on several in-house evaluation sets from different domains. Legend: y: yes; n: no, p: partly} \label{tab:results_on_other_evalsets}
		\begin{tabular}{|l|l|l|l|l||r|r|r|r|}
			\hline
			\textbf{Evaluation Set} & \textbf{Size}  & \textbf{Noise} & \textbf{Reverb}. & \textbf{Spont.} & \textbf{Baseline} & \textbf{Stage 2} &  \textbf{Stage 1} & \textbf{Two-}	\\
			& [min.] & & & \textbf{Speech} & & \textbf{only} & \textbf{only} & \textbf{Staged} \\
			\hline
			DiSCo Planned Clean	& 55	& n	& n	& n	&  9.03	& 9.23	&  8.95	& \textbf{8.89}\\ 
			DiSCo Spontaneous Clean	& 115	& n	& n	& y	&  10.25	& 10.06	&  \textbf{9.90}	& 9.94\\ 
			DiSCo Planned Mix	& 87	& y	& n	& n	&  11.64	& 11.67	&  \textbf{10.80}	& 10.83\\ 
			DiSCo Spontaneous Mix	& 66	& y	& n	& y	&  19.48	& 18.80	&  17.54	& \textbf{17.41}\\ 
			General German Broadcasts	& 61	& p	& p	& p	&  12.31	& 11.87	&  11.49	& \textbf{11.24}\\ 
			Challenging Broadcast Radio	& 52	& y	& p	& y	&  23.43	& 23.20	&  22.69	& \textbf{22.02}\\ 
			Challenging Broadcast TV	& 53	& y	& p	& y	&  17.78	& 17.44	&  17.28	& \textbf{17.02}\\ 
			Spoken QALD-7 & 15	& p	& y	& p	&  20.59	& 19.70	&  18.34	& \textbf{17.72}\\ 
			Humanities (Interaction)	& 49	& y	& y	& y	&  66.50	& 64.37	&  47.81	& \textbf{47.13}\\ 
			\hline
		\end{tabular}
	\end{center}
\end{table*}

\subsection{Leave-one-speaker-out experiments}
\begin{figure}
	\centering
	\centerline{\epsfig{figure=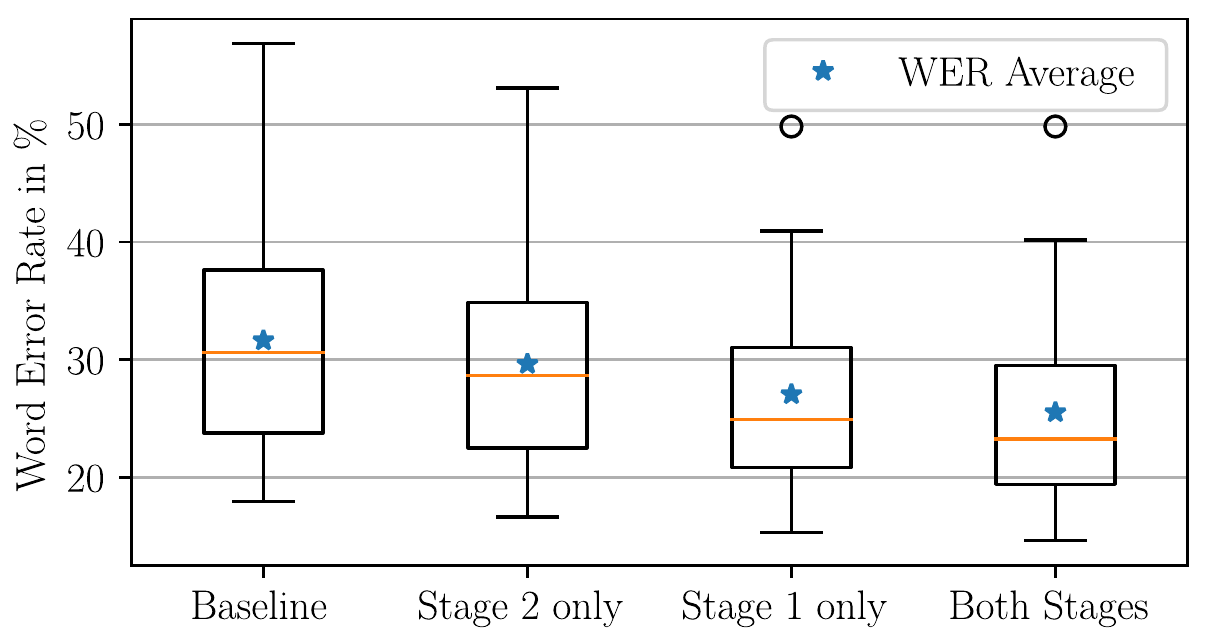,width=8.5cm}}
	\caption{Boxplot diagrams of the achieved WER in the 35 leave-one-speaker-out experiments for each setup. The star within the boxplots marks the average word error rate of all experiments w.r.t to the number of words in each of the sets. }
	\label{fig:boxplot}
\end{figure}

The results of the $ 35 $ leave-one-speaker-out experiments for the four different setups are given in form of a boxplot diagram in Fig.\ \ref{fig:boxplot}. Our experiment shows that the word error rates (WER) significantly decreases when applying the proposed approach, compared to the baseline. In the clean setup, only one half of the experiments achieve a WER below $ 30 \% $. However, in the proposed approach, this is the case for about $ 75 \% $ of the experiments. With the exception of one outlier, all experiments in the proposed approach have a WER below or near $ 40 \% $. Half of the experiments achieve a WER below $ 24 \% $ in the proposed approach. On average, the WER decreases from $ 31.6 \% $ in the clean setup to $ 25.5 \% $ using the two-staged approach.

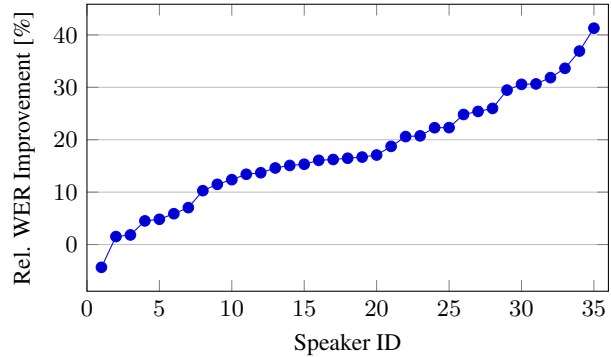
\begin{figure}
	\centering
		\begin{tikzpicture} 
		\begin{axis}[ 
		ymajorgrids,
		height=5.4cm,
		width=0.99\linewidth,
		style={font=\small},
		xlabel={Speaker ID},  
		ylabel={Rel. WER Improvement [\%]}, 
		xtick={0,5,...,35},
		ytick={-10,0,...,40},
		xmin=0,
		xmax=36,
		]
		
		\addplot coordinates {
		(1, -4.35)
		(2, 1.52)
		(3, 1.83)
		(4, 4.51)
		(5, 4.82)
		(6, 5.88)
		(7, 7.04)
		(8, 10.28)
		(9, 11.49)
		(10, 12.38)
		(11, 13.41)
		(12, 13.70)
		(13, 14.61)
		(14, 15.09)
		(15, 15.32)
		(16, 16.07)
		(17, 16.23)
		(18, 16.47)
		(19, 16.69)
		(20, 17.08)
		(21, 18.73)
		(22, 20.60)
		(23, 20.74)
		(24, 22.29)
		(25, 22.31)
		(26, 24.82)
		(27, 25.40)
		(28, 25.98)
		(29, 29.47)
		(30, 30.57)
		(31, 30.65)
		(32, 31.85)
		(33, 33.63)
		(34, 36.92)
		(35, 41.29)
		}; 
		
		\end{axis} 
		\end{tikzpicture}
	\caption{Relative word error rate improvement of the proposed approach compared to the clean baseline for each leave-one-speaker-out experiment. IDs are sorted by the improvement for better visualization.}
	\label{fig:relative_wer_improvement}
\end{figure}

The relative WER improvements of each leave-one-speaker-out experiment using the proposed approach compared to the clean-trained baseline model are shown in Fig.\ \ref{fig:relative_wer_improvement}. For $ 34 $ out of the $ 35 $ experiments the WER does decrease and only for one experiment the WER slightly increases. The speaker in this one experiment is recorded in a rather clean acoustic condition and has no noteworthy peculiarities in the nature of his speaking. For $ 27 $ out of the $ 35 $ experiments, the WER improves by more than $ 10 \% $ relative to the baseline.

The results using Stage 1 only are slightly worse than in the two-staged approach. On average, the WER achieved using only Stage 1 is $ 27.1 \% $. This means removing Stage 2 decreases the speech recognition performance by $ 6.3 \% $ relative. Removing Stage 1 and applying the transfer learning stage on the clean baseline model gives an average WER of $ 29.6 \% $. Thus, in the setup examined, transfer learning on a clean source model yields on average slightly worse results than the sole data augmentation in Stage 1---but is also a significant improvement to the baseline.

\begin{figure}[b!]
	\centering
	\begin{tikzpicture} 
	\begin{axis}[ 
	ymajorgrids,
	height=5.8cm,
	width=0.99\linewidth,
	style={font=\small},
	xlabel={Speaker ID},  
	ylabel={Rel.  WER Improvement [\%]}, 
	xtick={0,5,...,35},
	ytick={10,0,...,-60},
	xmin=0,
	xmax=36,
	legend style={at={(0.42,-0.1)},anchor=north,draw=none,font=\small},
	legend columns=2
	]
	
	\addplot coordinates {
		(1, 7.10)
		(2, 0.00)
		(3, 6.32)
		(4, 1.71)
		(5, 1.28)
		(6, -5.66)
		(7, 0.00)
		(8, -2.26)
		(9, -14.16)
		(10, -6.54)
		(11, -2.46)
		(12, -12.78)
		(13, -9.92)
		(14, -8.09)
		(15, -14.56)
		(16, -6.67)
		(17, -15.48)
		(18, -14.48)
		(19, -17.71)
		(20, -21.59)
		(21, -14.08)
		(22, -21.77)
		(23, -20.93)
		(24, -12.26)
		(25, -13.91)
		(26, -27.77)
		(27, -25.80)
		(28, -27.20)
		(29, -29.09)
		(30, -33.88)
		(31, -48.28)
		(32, -32.95)
		(33, -32.77)
		(34, -41.89)
		(35, -51.63)
	}; 
	\addlegendentry{Removed Data Augmen.}
	
	\addplot coordinates {
		(1, -10.63)
		(2, -3.89)
		(3, -6.27)
		(4, -3.42)
		(5, -6.35)
		(6, -6.78)
		(7, -2.53)
		(8, -5.93)
		(9, -9.40)
		(10, 0.00)
		(11, -9.88)
		(12, -0.45)
		(13, -8.97)
		(14, -7.31)
		(15, -5.74)
		(16, -10.39)
		(17, -8.54)
		(18, -8.55)
		(19, -7.32)
		(20, -4.88)
		(21, -5.63)
		(22, -3.43)
		(23, -5.66)
		(24, -7.39)
		(25, -4.66)
		(26, -1.92)
		(27, -5.16)
		(28, -5.69)
		(29, -7.45)
		(30, -10.16)
		(31, -6.79)
		(32, -11.10)
		(33, -8.04)
		(34, -4.73)
		(35, -3.30)
	}; 
	\addlegendentry{Removed Transfer Learning}
	
	\end{axis} 
	\end{tikzpicture}
	\caption{Relative WER change for each leave-one-speaker-out experiments in case one of the stages is removed from the proposed approach. Speaker IDs are in the same order as in Fig.\ \ref{fig:relative_wer_improvement}. Negative values indicate an increased WER.}
	\label{fig:ablation_study}
\end{figure}
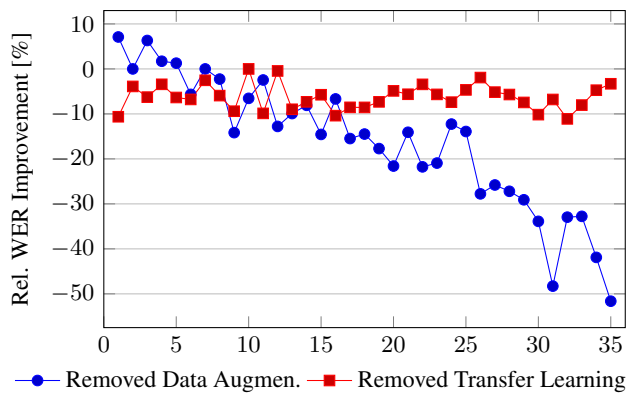

A more in-depth look at the $ 35 $ individual experiments is given in Fig.\ \ref{fig:ablation_study} where the relative WER improvement in comparison to the proposed two-staged approach is given when only one of the stages is applied. It is evident that the data augmentation has a large impact on the speech recognition performance in many of the experiments. However, for four experiments it even increases the WER. The improvement by the transfer learning on the other hand is quite consistent for the experiments.

\subsection{Robustness with several evaluation sets}
Finally, we investigate the robustness of the proposed approach by evaluating it on several different German in-house evaluation sets from other domains. Some of the sets partly share some challenges of oral history interviews, such as spontaneous speech or reverberation. 

DiSCo \cite{Baum.2010.DiSCoCorpus} is a corpus for the German broadcast domain and is split in four evaluation sets: planned and spontaneous speech each in clean and mixed acoustic conditions. The two Challenging Broadcast evaluation sets are similar to the DiSCo Spontaneous Mix set and contain several challenging interviews and recordings with a lot of spontaneous speech, often in challenging acoustic conditions, and even some overlapping speech. The Spoken QALD-7 corpus contains in-house recorded questions for a question answering system based on \cite{Usbeck.2017.promptsForSpeechAssistantEvalSet} by several speakers using a web interface and their respective microphone---a headset or build-in laptop microphone for instance. The in-house Humanities evaluation set contains recordings of people informally talking to each other about different topics recorded in challenging acoustic conditions.

For this experiment, we use the entire oral history set in the second stage for transfer learning and no data is held out for evaluation. The word error rates of such a model on the evaluation sets are given in Table \ref{tab:results_on_other_evalsets}. Even though we used the two-staged acoustic modeling adaption to improve the performance on oral history interviews, the model performs better than the comparison models on many of the evaluation sets. The increase in performance is higher on rather challenging test sets while maintaining or even slightly increasing the good performance on the more clean tasks. Therefore, we conclude that the two-staged approach not only is useful for a specific task but also helps to increase the generalization of the acoustic model.

\section{Conclusion}
In this work, we proposed a two-staged acoustic modeling adaption for robust speech recognition and evaluated the approach on the challenging example of German oral history interviews. We evaluated the reliability of our approach with a leave-one-speaker-out evaluation method in which we perform $ 35 $ experiments for one setup. We showed that the proposed approach increases the speech recognition performance in $ 34 $ of the $ 35 $ experiments and performs better than using one of the methods alone. On average, the word error rate decreases relatively by $ 19.3 \% $. Furthermore, we showed that our approach helps to increase the generalization of acoustic models and leads to increased recognition for challenging recordings while maintaining the good performance on clean tasks.

\section{Acknowledgements}
This research has been funded by the Federal Ministry of Education and Research of Germany (BMBF) in the project \textit{KA$^3 $ - Kölner Zentrum für Analyse und Archivierung von AV-Daten} (Cologne center for the analysis and archiving of audiovisual data) (project number: 01UG1811B).

\bibliographystyle{IEEEbib}
\bibliography{literature}

\begin{thebibliography}{10}

\bibitem{Povey.2016.LF_MMI}
Daniel Povey, Vijayaditya Peddinti, Daniel Galvez, Pegah Ghahremani, Vimal
  Manohar, Xingyu Na, Yiming Wang, and Sanjeev Khudanpur,
\newblock ``Purely sequence-trained neural networks for {ASR} based on
  lattice-free {MMI},''
\newblock in {\em 17th Annual Conference of the International Speech
  Communication Association (Interspeech)}, 2016, pp. 2751--2755.

\bibitem{Gref.2018.oralhistory01}
Michael Gref, Joachim K{\"{o}}hler, and Almut Leh,
\newblock ``Improved transcription and indexing of oral history interviews for
  digital humanities research,''
\newblock in {\em Eleventh International Conference on Language Resources and
  Evaluation ({LREC})}, 2018.

\bibitem{Gref.2018.oralhistory02}
Michael Gref, Christoph Schmidt, and Joachim K{\"o}hler,
\newblock ``Improving robust speech recognition for german oral history
  interviews using multi-condition training,''
\newblock in {\em 13. {ITG} Symposium on Speech Communication}. 2018, pp.
  256--260, {IEEE}.

\bibitem{Ko.2017.KaldiReverbDataAugmentation}
Tom Ko, Vijayaditya Peddinti, Daniel Povey, Michael~L. Seltzer, and Sanjeev
  Khudanpur,
\newblock ``A study on data augmentation of reverberant speech for robust
  speech recognition,''
\newblock in {\em {IEEE} International Conference on Acoustics, Speech and
  Signal Processing ({ICASSP})}, 2017, pp. 5220--5224.

\bibitem{Ko.2015.SpeedPerturbation}
Tom Ko, Vijayaditya Peddinti, Daniel Povey, and Sanjeev Khudanpur,
\newblock ``Audio augmentation for speech recognition,''
\newblock in {\em 16th Annual Conference of the International Speech
  Communication Association (Interspeech)}, 2015, pp. 3586--3589.

\bibitem{Goodfellow.2016.DeepLearningBook}
Ian~J. Goodfellow, Yoshua Bengio, and Aaron~C. Courville,
\newblock {\em Deep Learning},
\newblock Adaptive computation and machine learning. {MIT} Press, 2016,
\newblock pp. 526--528.

\bibitem{Wang.2015.transfer_learning_overview_in_speechprocessing}
Dong Wang and Thomas~Fang Zheng,
\newblock ``Transfer learning for speech and language processing,''
\newblock in {\em Asia-Pacific Signal and Information Processing Association
  Annual Summit and Conference ({APSIPA})}, 2015, pp. 1225--1237.

\bibitem{Ghahremani.2017.KaldiTransferlearning_LF_MMI}
Pegah Ghahremani, Vimal Manohar, Hossein Hadian, Daniel Povey, and Sanjeev
  Khudanpur,
\newblock ``Investigation of transfer learning for {ASR} using {LF-MMI} trained
  neural networks,''
\newblock in {\em {IEEE} Automatic Speech Recognition and Understanding
  Workshop ({ASRU})}, 2017, pp. 279--286.

\bibitem{Jeub.2009.AachenRirDB}
M.~Jeub, M.~Schafer, and P.~Vary,
\newblock ``A binaural room impulse response database for the evaluation of
  dereverberation algorithms,''
\newblock in {\em 16th International Conference on Digital Signal Processing},
  2009, pp. 1--5.

\bibitem{Cheng.2017.KaldiLSTMDropout}
Gaofeng Cheng, Vijayaditya Peddinti, Daniel Povey, Vimal Manohar, Sanjeev
  Khudanpur, and Yonghong Yan,
\newblock ``An exploration of dropout with lstms,''
\newblock in {\em 18th Annual Conference of the International Speech
  Communication Association (Interspeech)}, 2017, pp. 1586--1590.

\bibitem{Bisani.2008.SequiturG2P}
Maximilian Bisani and Hermann Ney,
\newblock ``Joint-sequence models for grapheme-to-phoneme conversion,''
\newblock {\em Speech Communication}, vol. 50, no. 5, pp. 434--451, 2008.

\bibitem{Stadtschnitzer.2014.GerTV1000hCorpus}
Michael Stadtschnitzer, Jochen Schwenninger, Daniel Stein, and Joachim
  K{\"{o}}hler,
\newblock ``Exploiting the large-scale german broadcast corpus to boost the
  fraunhofer {IAIS} speech recognition system,''
\newblock in {\em Ninth International Conference on Language Resources and
  Evaluation ({LREC})}, 2014, pp. 3887--3890.

\bibitem{Dehak.2011.ivector}
Najim Dehak, Patrick Kenny, R{\'{e}}da Dehak, Pierre Dumouchel, and Pierre
  Ouellet,
\newblock ``Front-end factor analysis for speaker verification,''
\newblock {\em {IEEE} Trans. Audio, Speech {\&} Language Processing}, vol. 19,
  no. 4, pp. 788--798, 2011.

\bibitem{Waibel.1989.TDNN}
Alexander~H. Waibel, Toshiyuki Hanazawa, Geoffrey~E. Hinton, Kiyohiro Shikano,
  and Kevin~J. Lang,
\newblock ``Phoneme recognition using time-delay neural networks,''
\newblock {\em {IEEE} Trans. Acoustics, Speech, and Signal Processing}, vol.
  37, no. 3, pp. 328--339, 1989.

\bibitem{Peddinti.2015.KaldiTDNN}
Vijayaditya Peddinti, Daniel Povey, and Sanjeev Khudanpur,
\newblock ``A time delay neural network architecture for efficient modeling of
  long temporal contexts,''
\newblock in {\em 16th Annual Conference of the International Speech
  Communication Association (Interspeech)}, 2015, pp. 3214--3218.

\bibitem{Hochreiter.1997.LSTM}
Sepp Hochreiter and J{\"{u}}rgen Schmidhuber,
\newblock ``Long short-term memory,''
\newblock {\em Neural Computation}, vol. 9, no. 8, pp. 1735--1780, 1997.

\bibitem{Gers.2000.LSTMLearningToForget}
Felix~A. Gers, J{\"{u}}rgen Schmidhuber, and Fred~A. Cummins,
\newblock ``Learning to forget: Continual prediction with {LSTM},''
\newblock {\em Neural Computation}, vol. 12, no. 10, pp. 2451--2471, 2000.

\bibitem{Gers.2000.LSTMPeepholes}
Felix~A. Gers and J{\"{u}}rgen Schmidhuber,
\newblock ``Recurrent nets that time and count,''
\newblock in {\em {IJCNN} {(3)}}, 2000, pp. 189--194.

\bibitem{Sak.2014.LSTMProjection}
Hasim Sak, Andrew~W. Senior, and Fran{\c{c}}oise Beaufays,
\newblock ``Long short-term memory recurrent neural network architectures for
  large scale acoustic modeling,''
\newblock in {\em 15th Annual Conference of the International Speech
  Communication Association (Interspeech)}, 2014, pp. 338--342.

\bibitem{Povey.2011.KaldiToolkit}
Daniel Povey, Arnab Ghoshal, Gilles Boulianne, Lukas Burget, Ondrej Glembek,
  Nagendra Goel, Mirko Hannemann, Petr Motlicek, Yanmin Qian, Petr Schwarz, Jan
  Silovsky, Georg Stemmer, and Karel Vesely,
\newblock ``The kaldi speech recognition toolkit,''
\newblock in {\em IEEE Workshop on Automatic Speech Recognition and
  Understanding ({ASRU})}. Dec. 2011, IEEE Signal Processing Society.

\bibitem{Baum.2010.DiSCoCorpus}
Doris Baum, Daniel Schneider, Rolf Bardeli, Jochen Schwenninger, Barbara
  Samlowski, Thomas Winkler, and Joachim K{\"{o}}hler,
\newblock ``{DiSCo} - {A} german evaluation corpus for challenging problems in
  the broadcast domain,''
\newblock in {\em International Conference on Language Resources and Evaluation
  ({LREC})}, 2010.

\bibitem{Usbeck.2017.promptsForSpeechAssistantEvalSet}
Ricardo Usbeck, Axel{-}Cyrille~Ngonga Ngomo, Bastian Haarmann, Anastasia
  Krithara, Michael R{\"{o}}der, and Giulio Napolitano,
\newblock ``7th open challenge on question answering over linked data
  {(QALD-7)},''
\newblock in {\em Semantic Web Challenges - 4th SemWebEval Challenge at
  ({ESWC})}, 2017, pp. 59--69.

\end{thebibliography}

\end{document}